\documentclass[preprint]{revtex4}
\usepackage{amssymb}
\usepackage{graphicx}

\begin{document}

\title{Statistics of mixing in three-dimensional Rayleigh--Taylor 
turbulence at low Atwood number and Prandtl number one}

\author{G. Boffetta$^{1}$, A. Mazzino$^{2}$, S. Musacchio$^{3}$  
and L. Vozella$^{2}$}
\affiliation{$^1$Dipartimento di Fisica Generale and INFN, Universit\`a di Torino,
via P.Giuria 1, 10125 Torino (Italy) \\
and CNR-ISAC, Sezione di Torino, corso Fiume 4, 10133 Torino (Italy)}
\affiliation{$^2$Dipartimento di Fisica, Universit\`a di Genova, INFN and CNISM,
via Dodecaneso 33, 16146 Genova (Italy)}

\affiliation{$^3$ CNRS UMR 6621, Lab. J.A. Dieudonnn\'e, Universit\'e de Nice Sophia-Antipolis, Parc Valrose, 06108 Nice Cedex 02 (France)}
\date{\today}

\begin{abstract}
Three-dimensional miscible Rayleigh--Taylor (RT) turbulence at 
small Atwood number and at Prandtl number one is investigated by 
means of high resolution direct numerical simulations of the 
Boussinesq equations.
RT turbulence is a paradigmatic time-dependent turbulent system 
in which the integral scale grows in time following the evolution
of the mixing region.
In order to fully characterize the statistical properties 
of the flow, both temporal and spatial behavior of relevant 
statistical indicators have been analyzed.  

Scaling of both global quantities ({\it e.g.}, Rayleigh, Nusselt and 
Reynolds numbers) and scale dependent observables built in terms of 
velocity and temperature fluctuations are considered. 
We extend the mean-field analysis for velocity and temperature 
fluctuations to take into account intermittency, both in time and 
space domains. 
We show that the resulting scaling exponents are compatible with 
those of classical Navier--Stokes turbulence advecting a passive 
scalar at comparable Reynolds number. Our  
results support the scenario of universality of turbulence with respect 
to both the injection mechanism and the geometry of the flow. 
\end{abstract}


\maketitle

\section{Introduction}
\label{sec:1}

The Rayleigh--Taylor (RT) instability is a well-known fluid-mixing 
mechanism originating at the
interface between a light fluid accelerated into an heavy fluid.
It was first described by Rayleigh \cite{rayleigh_plms83} 
for incompressible fluid under gravity and later generalized
to all accelerated fluid by Taylor \cite{taylor_prsl50}.

RT instability plays a crucial role in many 
fields of science and technology. In particular, in gravitational
fusion it has been recognized as the dominant acceleration mechanism 
for thermonuclear reactions in type-Ia supernovae \cite{cc_natphys06,zingale05}.
The efficiency of inertial confinement fusion depends dramatically on
the ability to suppress RT instability on the interface between the fuel and
the pusher shell \cite{regan02,fujioka_etal04}. 

In a late stage, RT instability develops 
into the so-called RT turbulence in which
a layer of mixed fluid grows in time increasing the
kinetic energy of the flow at the expenses of the potential 
energy. 
This process finds applications in many fields, e.g. atmospheric
and oceanic buoyancy driven mixing.
Despite the great importance and long history of 
RT turbulence, a consistent phenomenological theory has been
proposed only recently \cite{chertkov_prl03}.
In three dimensions, this theory predicts a Kolmogorov-like scenario,
with a quasi-stationary energy cascade in the mixing layer. 
The prediction is based on the Kolmogorov--Obukhov picture of
turbulence in which density fluctuations are transported passively 
in the cascade and kinetic-energy flux is scale independent 
\cite{frisch_95}.
Quasi-stationarity is a consequence of 
Kolmogorov scaling of characteristic times associated to turbulent eddies: 
large-scales grow driven from potential 
energy, while small-scale structures, fed by the turbulent cascade, 
follow adiabatically large-scale growth. 
These theoretical predictions have been partially confirmed 
by recent numerical studies \cite{cc_natphys06,vc_pof09,matsumoto,bmmv_pre09}.
Other alternative phenomenological approaches 
(see {\it e.g.} \cite{Zhou}) does
not necessarily lead to the Kolmogorov scaling for the energy spectra.

In this Paper we carry out an analysis of 
the scaling behavior of relevant observables with the aim of  
 deepening our previous investigation \cite{bmmv_pre09}.  
Indeed, our aim is to make a careful investigation of  
the time evolution of global observables and of spatial/temporal 
scaling and intermittency. 
Moreover we push the analogy of RT turbulence with 
usual Navier--Stokes (NS) turbulence much further.
We show that small-scale velocity and temperature fluctuations develop intermittent 
distributions with structure-function scaling exponents
consistent with NS turbulence advecting a passive scalar. 

This Paper is organized as follows. In Sec.~\ref{sec:2} we formulate 
the problem and outline the phenomenology. 
After providing a description of the numerical 
setup in Sec.~\ref{sec:3}, we describe our results in the subsequent Sections.  
Sec.~\ref{sec:3_1} is devoted to the investigation of the temporal 
evolution of global quantities. 
In Sec.~\ref{sec:4} we focus on the statistics at small scales. 
Finally, the Conclusions are provided by 
summarizing the main results. 

\section{Equation of motion and phenomenology}
\label{sec:2}

We consider the three-dimensional Boussinesq equations for an incompressible
velocity field (${\bf \nabla} \cdot {\bf v}=0$), 
\begin{eqnarray}
&& \partial_t {\bf v} + {\bf v} \cdot {\bf \nabla} {\bf v} = - {\bf \nabla} p
+ \nu \triangle {\bf v} - \beta {\bf g} T \label{eq:1} \\
&& \partial_t T + {\bf v} \cdot {\bf \nabla} T = \kappa \triangle T 
\label{eq:2}
\end{eqnarray}
$T({\bf x},t)$ being the temperature field, proportional to
the density via the thermal expansion coefficient $\beta$ as 
$\rho=\rho_0 [1-\beta (T-T_0)]$ ($\rho_0$ and $T_0$ are reference
values),
$\nu$ is the kinematic viscosity, $\kappa$ the molecular
diffusivity and ${\bf g}=(0,0,-g)$ the gravitational acceleration.

At time $t=0$ the system is at rest with cooler (heavier, density $\rho_2$) 
fluid placed above the hotter (lighter, density $\rho_1$) one. 
This corresponds to ${\bf v}({\bf x},0)=(0,0,0)$ and
to a step function for the initial temperature profile: 
$T({\bf x},0)=-(\theta_0/2) \mbox{sgn}(z)$ where $\theta_0$ 
is the temperature jump which fixes the 
Atwood number $A=(\rho_2-\rho_1)/(\rho_2+\rho_2)=(1/2) \beta \theta_0$.
The development of the instability leads to a mixing zone
of width $h$ which starts from the plane $z=0$ and is dimensionally
expected to grow in time according to 
$h(t)= \alpha A g t^2$ (where $\alpha$ is a dimensionless constant 
to be determined) which implies the relation $v_{rms} \simeq A g t$ 
for typical velocity fluctuations (root mean square velocity)
inside the mixing zone.

The convective state is characterized by the turbulent heat flux and
energy transfer as a function of mean temperature gradient. In terms
of dimensionless variables these quantities are represented respectively
by the Nusselt number $Nu=1+\langle w T \rangle h/(\kappa \theta_0)$
($w$ being the vertical velocity) 
and the Reynolds number $Re=v_{rms} h/\nu$ as a function of 
the Rayleigh number $Ra=\beta g \theta_0 h^3/(\nu \kappa)$ and the 
Prandtl number $Pr=\nu/\kappa$. 
Here and in the following $\langle ... \rangle$
denotes spatial average inside the turbulent mixing zone, while the overbar
indicates the average over horizontal planes at fixed $z$.

One of the most important problems in thermal convection is to 
find the functional
relation between the convective state characterized by $Nu$ and $Re$
and the parameter space defined by 
$Ra$ and $Pr$ \cite{siggia_arfm94}.
The existence of an asymptotic regime at high $Ra$, with a simple
power law dependence $Nu \sim Ra^{\xi}$ and $Re \sim Ra^{\gamma}$,
is still controversial in the case of Rayleigh--B\'enard convection,
despite the number of experiments at very large $Ra$.
Most of the experiments have reported an exponent $\xi \simeq 0.3$
\cite{gsns_nat99,nssd_nat00} of a more complex behavior
\cite{xba_prl00,na_prl03} partially described by a phenomenological
theory \cite{gl_jfm00}.
However, many years ago, Kraichnan \cite{kraichnan_pof62}
predicted an asymptotic exponent $\xi=1/2$ (with logarithmic 
corrections) associated to the now called ``ultimate state of thermal 
convection'', while exponents $\xi > 1/2$ are excluded by a
rigorous upper bound $Nu \le (1/6) Ra^{1/2}-1$ \cite{dc_pre96}.
The ultimate state regime is expected
to hold when thermal and kinetic boundary layers become irrelevant,
and indeed has been observed in numerical simulations of thermal
convection at moderate $Ra$ when boundaries are removed \cite{lt_prl03},
while no indication of ultimate state regime has been observed
in Rayleigh--B\'enard experiments \cite{gsns_nat99}.

The ultimate state exponent is formally derived from 
kinetic energy and temperature balance equations \cite{gl_jfm00}.
In the present context of RT turbulence they can
more easily be obtained from the temporal scaling of $h$ and $v_{rms}$.
Assuming that $\langle wT \rangle \sim v_{rms} \theta_0$,
using the above definitions one estimates:
\begin{equation}\label{eqnutemp}
 Ra \simeq (Ag)^4 t^6/(\nu \kappa), \quad
Re \simeq (Ag)^2 t^3/\nu \quad \textrm{and} \quad Nu \simeq (Ag)^2 t^3/\kappa 
\end{equation}
from which
\begin{equation}
Nu \sim Pr^{1/2} Ra^{1/2} \qquad \textrm{and} \qquad 
Re \sim Pr^{-1/2} Ra^{1/2}
\label{eq:3}
\end{equation}

For what concerns the small-scale statistics inside the
mixing zone, the phenomenological theory \cite{chertkov_prl03}
predicts for the 3D case an adiabatic Kolmogorov--Obukhov scenario with a 
time-dependent kinetic-energy flux 
$\epsilon \simeq v_{rms}^3/h \simeq (\beta g \theta_0)^2 t$.
Spatial-temporal scaling of velocity and temperature fluctuations
are therefore expected to follow
\begin{eqnarray}
\delta_r v(t) &\simeq & \epsilon^{1/3} r^{1/3} \simeq (\beta g \theta_0)^{2/3} t^{1/3} r^{1/3}
\label{eq:4} \\
\delta_r T(t) &\simeq & \epsilon^{-1/6}\epsilon_T^{1/2}r^{1/3} \simeq \theta_0^{2/3} (\beta g)^{-1/3} t^{-2/3} r^{1/3}
\label{eq:5}
\end{eqnarray}
where $\delta_r v(t)=v(x+r,t)-v(x,t)$ is the velocity increment on
a separation $r$ (similarly for temperature) and
$\epsilon_T\simeq {\theta_0}^2 t^{-1}$ is the temperature-variance flux. 
We remark that the above scaling is consistent with
the assumption of the theory that temperature fluctuations are
passively transported at small scales (indeed using (\ref{eq:4}-\ref{eq:5})
the buoyancy term $\beta g T$
becomes subleading in (\ref{eq:1}) at small scales). This is the
main difference with respect to the 2D case in which
temperature fluctuations force the turbulent flow at all scales
\cite{chertkov_prl03,cmv_prl06, zingale05}.

\section{Numerical setting}
\label{sec:3}

The Boussinesq equations (\ref{eq:1}-\ref{eq:2}) are integrated
by a standard $2/3$-dealiased pseudospectral method on a 
three-dimensional periodic domain of square basis $L_x=L_y$ and aspect
ratio $L_x/L_z=R$ with uniform grid spacing at different resolutions
as shown in Table~\ref{table1}.
In the following, all physical quantities are made dimensionless using
the vertical scale $L_z$, the temperature jump $\theta_0$ and the
characteristic time $\tau=(L_z/A g)^{1/2}$ as fundamental units.

\begin{table}
\begin{tabular}{c|cccccc}
Label & $N_x=N_y$ & $N_z$ & $\nu=\kappa$ & $R_{\lambda}$ \\ \hline
A & $256$ & $1024$ & $9.5 \times 10^{-6}$ & 103 \\
B & $512$ & $2048$ & $4.8 \times 10^{-6}$ & 196  \\
C & $1024$ & $1024$ & $3.2 \times 10^{-6}$ & 122 \\
\end{tabular}
\caption{Parameters of the simulations. $N_x$, $N_y$, $N_z$
spatial resolution, $\nu$ viscosity, 
$\kappa$ thermal diffusivity,
$R_{\lambda}=v_{rms}^2\sqrt{15/(\nu \epsilon)}$
Reynolds number evaluated at the end of the simulation.
All dimensional quantities are made dimensionless using the 
vertical box size $L_z$, the characteristic time $\tau=(L_z/A g)^{1/2}$
and the temperature jump $\theta_0$ as reference units.} 
\label{table1}
\end{table}

Time evolution is obtained by a second-order Runge--Kutta scheme 
with explicit linear part. In all the runs, $\beta g=2.0$ and
$Pr=\nu/\kappa=1$. Viscosity is sufficiently large
to resolve small scales ($k_{max} \eta \simeq 1.2$ at final time, 
being $\eta = \nu^{3/4}\epsilon^{-1/4}$ the Kolmogorov scale and 
$k_{max}=N_x/3$).

\begin{figure}[htb!]
\includegraphics[clip=true,keepaspectratio,width=12.0cm]{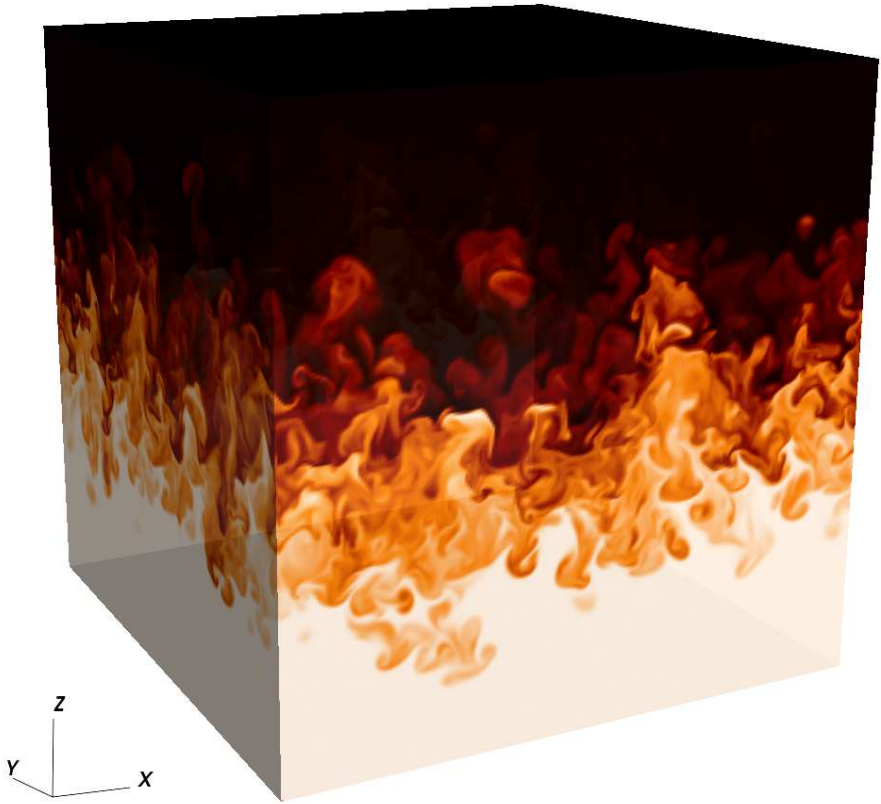}
\caption{Snapshot of temperature field for Rayleigh-Taylor simulation 
at $t=2 \tau$.  White (black) regions corresponds to hot (cold) fluid.
Parameters in Table~\ref{table1}, run B.}
\label{fig3.0}
\end{figure}

RT instability is seeded by perturbing the initial
condition with respect to the unstable step profile. 
Two different perturbations were implemented in order 
to check the independence of the turbulent
state from initial conditions. 
In the first case the interface $T=0$ at $z=0$ is perturbed by a 
superposition of two-dimensional waves of small amplitude $h_0=0.004 L_z$
in an isotropic range of wavenumbers $32 \le k \le 64$ (with $k^2=k_x^2+k_y^2$)
and random phases \cite{rda_jfm05}.
For the second set of simulations, we perturbed
the initial condition by adding $10 \%$ of white noise to the value
of $T({\bf x},0)$ in a layer of width $h_0$ around $z=0$.
Figure~\ref{fig3.0} shows a snapshot of the temperature field
in a cubic slice around $z=0$ in the turbulent regime at time $t=2 \tau$
for simulation $B$ (see Table~\ref{table1}).

\section{Evolution of global quantities}
\label{sec:3_1}
\begin{figure}[htb!]
\includegraphics[clip=true,keepaspectratio,width=12.0cm]{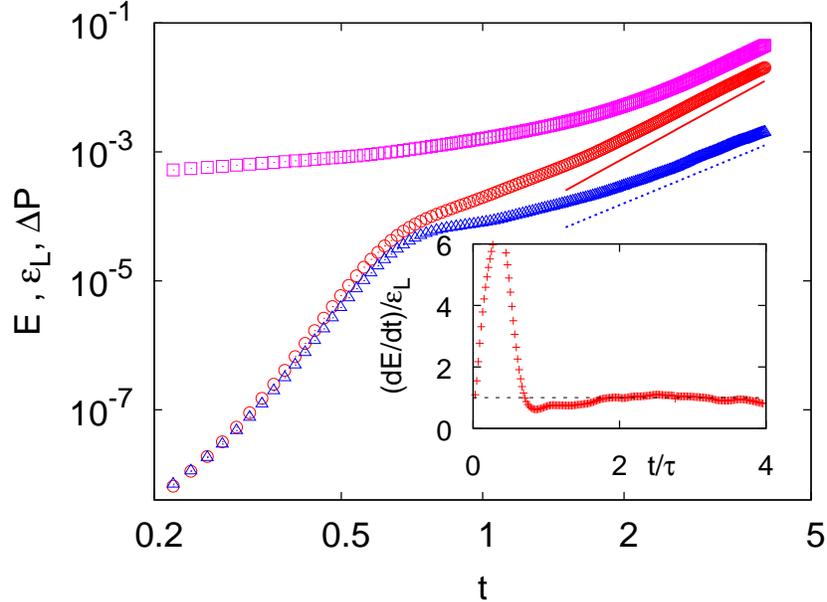}
\caption{Temporal growth of kinetic energy $E$ (red circles), 
kinetic-energy dissipation $\epsilon_L$ (blue triangles) and 
potential-energy loss $\Delta P$ (pink squares) for run $B$.
For clarity of the plot $\epsilon_L$ has been shifted by a factor $10$.
The two short lines represent the dimensional 
scaling $E(t) \sim \Delta P(t) \sim t^4$ 
and $\epsilon_L(t) \sim t^3$. Inset: 
ratio of the energy growth rate $dE/dt$ and the flux $\epsilon_L$.  
Data from run B.
}
\label{fig3.1}
\end{figure}

Figure~\ref{fig3.1} displays the evolution of the total kinetic 
energy $E=\int (1/2) v({\bf x})^2  \,d{\bf x}$ and total 
kinetic-energy dissipation $\epsilon_L$ as a function of time. 
After the linear instability regime, at $t \simeq \tau$ 
the turbulent regime sets in with algebraic time dependence. 
Temporal evolution of the two quantities are easily obtained recalling
that, being global quantities, an additional geometrical factor 
$h(t) \sim t^2$ due to the integration over the vertical direction 
has to be included. Therefore the predictions
are $E(t) \sim v_{rms}^2 h \sim t^4$ and 
$\epsilon_L \sim \epsilon h \sim t^3$, 
as indeed observed at late times. 
We also plot in Fig.~\ref{fig3.1} the total potential-energy
loss, defined as $P(0)-P(t)$ with $P(t)=-\beta g \int z\,T({\bf x})\,d{\bf x}$
which has the same temporal scaling of $E(t)$ as it is obvious from
energy balance: $d(E + P)/dt = -\epsilon_L$.
Notice that for this non-stationary turbulence the energy balance 
does not fix the ratio between 
the energy growth rate $dE/dt$ and the energy dissipation (and flux)
$\epsilon_L$. In the turbulent regime, our simulations show an 
``equipartition'' between large-scale energy growth and small-scale
energy dissipation:  $dE/dt \simeq \epsilon_L \simeq -(1/2)\,dP/dt$. 
This amounts to saying that  half of the power injected into 
the flow contributes to the growth 
of the large-scale flow, and half feeds the turbulent cascade
(see inset of Fig.~\ref{fig3.1}). This result was found to be
independent on the value of viscosity (the only adjustable parameter in the
system) and is consistent with previous findings 
\cite{ra_jfm04}.

An interesting remark is that
RT turbulence represents an instance of the general case of a turbulent
flow adiabatically evolving under a time-dependent energy input 
density $\mathcal{I}(t)$ 
which forces the flow at the integral scale $L(t)$ 
(concerning the problem of turbulent flow characterized by 
a time dependent forcing see, for example, 
\cite{lohse_pre03,lohse_cf08} and references therein). 
Energy balance requires
$d\mathcal{E}/dt=\mathcal{I}(t)-\epsilon(t)$, 
where $\mathcal{E}$ is the kinetic energy density.
Assuming a Kolmogorov spectrum for 
velocity fluctuations at scales smaller than the integral scale,
one estimates $\mathcal{E}(t) \simeq \epsilon^{2/3} L^{2/3}$.
Therefore, in situations characterized by an algebraic growth of the 
energy input density $\mathcal{I}(t) \sim t^{\gamma}$
a self-similar evolution of the energy spectrum
can be obtained only if  
$\epsilon(t) \sim t^{\gamma}$
and 
$L(t) \sim t^{(3+\gamma)/2}$. 
This is indeed realized in RT
turbulence, where $\gamma = 1$ and $\epsilon \sim t, L(t) \sim t^2$.   

\begin{figure}[htb!]
\includegraphics[clip=true,keepaspectratio,width=12.0cm]{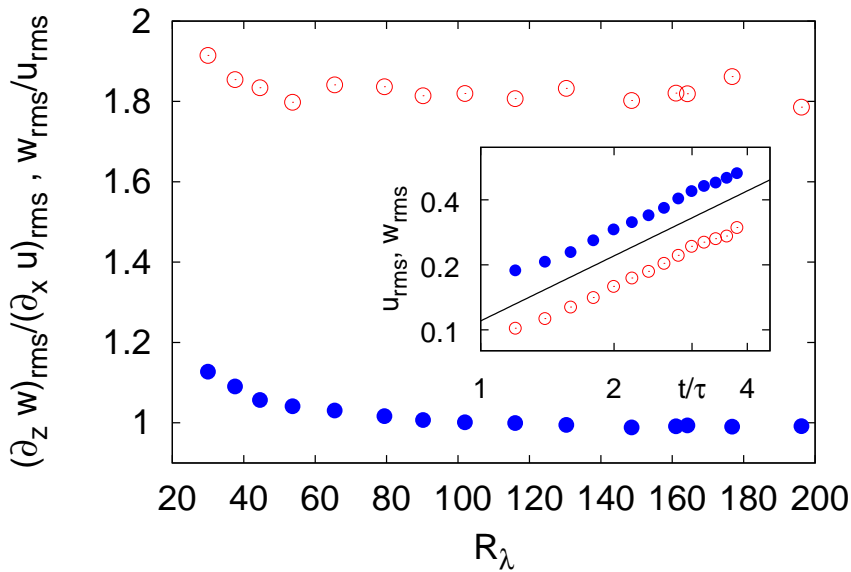}
\caption{Ratio of the vertical rms velocity $w_{rms}$ to the 
horizontal rms velocity $u_{rms}$ (red open circles) and
ratio of the vertical velocity gradient $(\partial_z w)_{rms}$ 
to the horizontal velocity gradient $(\partial_x u)_{rms}$ (blue
filled circles)
versus Reynolds number $R_{\lambda}$
indicating the recovery of isotropy at small scales. 
Inset: 
temporal evolution of horizontal rms velocity $u_{rms}$ (red open circles) 
and vertical rms velocity $w_{rms}$ (blue filled circles). 
The black line represents linear scaling. Data from run B.}
\label{fig3.2}
\end{figure}

In the inset of Fig.~\ref{fig3.2} the growth 
of vertical and horizontal rms velocity ($w_{rms}$ and $u_{rms}$ respectively), 
computed within the mixing layer, is shown. Both $u_{rms}$ and $w_{rms}$ grow linearly in
time, as expected, with the vertical velocity about twice the
horizontal one, reflecting the anisotropy of the forcing due to 
gravity.
It is interesting to observe that anisotropy decays at small scales,
where almost complete isotropy is recovered, as shown in Fig.~\ref{fig3.2}.
The ratio of vertical to horizontal rms velocity reaches a value
$w_{rms}/u_{rms} \simeq 1.8$ at later times 
(corresponding to 
$R_{\lambda} \simeq 200$)
while for the gradients we have
$(\partial_z w)_{rms}/(\partial_x u)_{rms} \simeq 1.0$. 

\begin{figure}[htb!]
\includegraphics[clip=true,keepaspectratio,width=12.0cm]{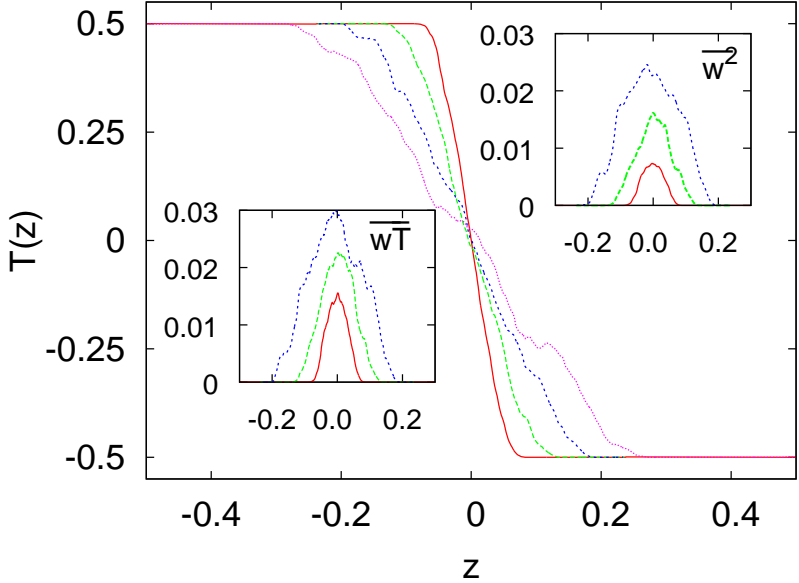}
\caption{Mean temperature profiles $\overline{T}(z,t)$ for a single realization
of simulation B with diffused initial perturbation
at times $t=1.4 \tau$, $t=2.0 \tau$, $t=2.6 \tau$ and $t=3.2 \tau$.
Lower and upper insets: profiles of the heat flux $\overline{w T}(z,t)$ 
and square vertical velocity $\overline{w^2}(z,t)$ at times
$t=1.4 \tau$, $t=2.0 \tau$ and $t=2.6 \tau$.}
\label{fig3.3}
\end{figure}

The evolution of the mean temperature profile 
$\bar{T}(z,t)\equiv 1/(L_x L_y) \int T({\bf x},t) dx dy $
is shown in Fig.~\ref{fig3.3}.
As observed in previous simulations \cite{cd_jfm01,cmv_prl06, matsumoto,bmmv_pre09}
the mean profile is approximately linear within the mixing
layer (where therefore the system recovers statistical 
homogeneity). 
Nevertheless, statistical fluctuations of temperature 
in the mixing layer are relatively strong: at later time we find a
flat profile of fluctuations. 
Moreover their distribution is close to a Gaussian 
with a standard deviation $\sigma_{T}(z) \simeq 0.25 \theta_0$ (not shown here).

In Fig.~\ref{fig3.3} we also plot the profile of the heat flux 
$\overline{w T}(z, t)$ and the square vertical velocity $\overline{w^2}(z, t)$.
Both vanish outside the mixing layer and inside show a similar shape not far
from a parabola. Of course, the time behaviors  
of the heat-flux and of the square vertical  velocity amplitude are  
different. Indeed, the former is expected to grow as $\propto t$ and the 
latter as $\propto t^2$. 

The mean temperature profile defines the width of the mixing layer.
Different definitions of the mixing width, $h$, have been proposed on 
the basis of integral
quantities or threshold values (see \cite{dly_jfm99} for a discussion
of the different methods). In the following we will use the simple
definition based on a threshold value: $\bar{T}(\pm h/2)=s \theta_0/2$
where $s<1$ represents the threshold. 

\begin{figure}[htb!]
\includegraphics[clip=true,keepaspectratio,width=12.0cm]{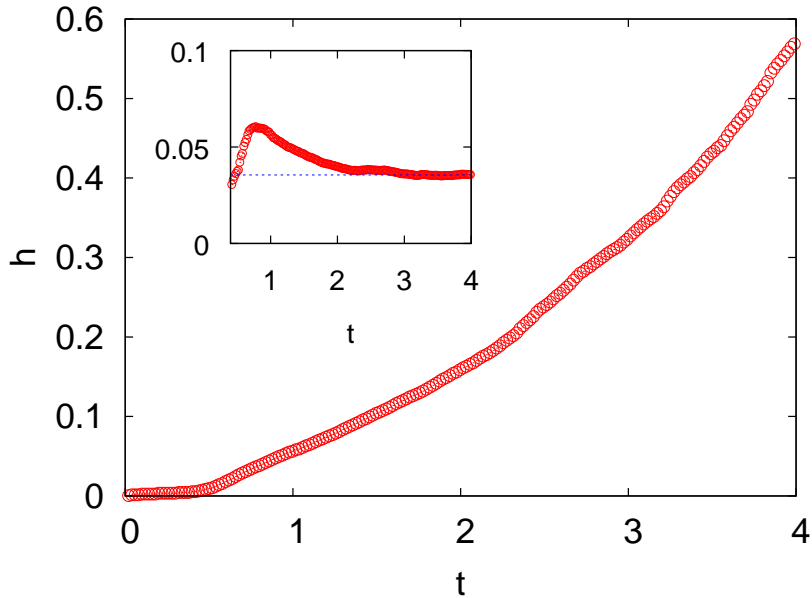}
\caption{Evolution of the mixing-layer width $h$ as a function of 
time $t$ for simulation B computed from 
the profiles of Fig.~\ref{fig3.3} with a threshold $s=0.8$.
The inset shows the compensation with dimensional prediction
$h/(A g t^2)$ converging to a value $\simeq 0.036$. 
}
\label{fig3.4}
\end{figure}

The evolution of the mixing width for $s=0.8$ is shown in Fig.~\ref{fig3.4}.
After an initial stage ($t<0.3\,\tau$) in which the perturbation 
relaxes towards the most unstable direction, we observe a
short exponential growth corresponding to the linear RT instability.
At later times ($t>0.6\,\tau$) the similarity regime sets in and 
the dimensional $t^2$ law is observed. 
The na\"ive compensation with $A g t^2$ gives an
asymptotic constant value $h/(A g t^2) \simeq 0.036$ 
for $t \ge 3 \,\tau$ and $Re\simeq 10^{4}$ 
(at which the mixing width is still below half box). 
For the calculation of $\alpha$,  more sophisticated analysis
have been proposed recently \cite{cc_natphys06,rc_jfm04,ccm_jfm04}.
using slightly different approaches (briefly, in \cite{rc_jfm04} a 
similarity assumption and in \cite{ccm_jfm04}
a mass flux and energy balance argument).
In both cases, the authors derive for the evolution of $h(t)$ the equation
\begin{equation}
\dot{h}^{2}=4\alpha A g h 
\label{eq:mona}
\end{equation}
which has solution
$h(t)= \alpha A g t^2 + 2 (\alpha A h_0)^{1/2} t + h_0$
where $h_0$ is the initial width introduced by the perturbation.
$\alpha=\dot{h}^2 /(4 A g h)$.
The idea is to get rid of the subleading terms and extract the $t^2$ 
contribution at early time by using directly (\ref{eq:mona}) 
and evaluating $\alpha=\dot{h}^2 /(4 A g h)$.

\begin{figure}[htb!]
\includegraphics[clip=true,keepaspectratio,width=12.0cm]{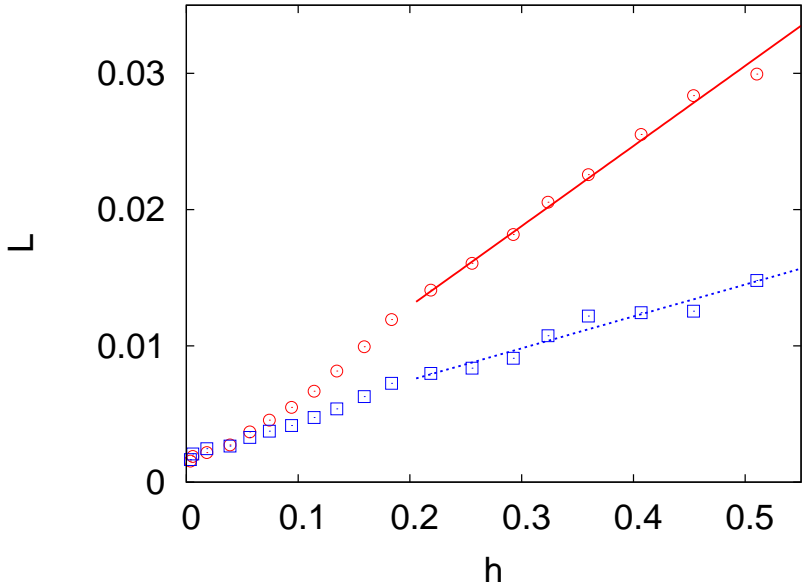}
\caption{Growth of the integral scale $L$ based on the vertical velocity
(red circles), and horizontal velocities (blue squares) 
as a function of the mixing layer $h$. Data from simulation B.}
\label{fig3.5}
\end{figure}

The growth of the mixing layer width $h(t)$, a geometrical quantity, 
is accompanied by the growth of the integral
scale $L(t)$, a dynamical quantity representing the typical size 
of the large-scale turbulent
eddies. Following Ref.~\cite{vc_pof09} we define $L$ as the half width of the
velocity correlation function 
$f(L) = \langle v_i(r) v_i(r+L) \rangle / \langle v^2 \rangle = 1/2$. 
In the turbulent regime the integral scale and the mixing length are linearly
related (see Fig~\ref{fig3.5}). A linear fit gives  
$L/h \simeq 1/17$ and $L/h \simeq 1/42$ for the integral
scale based on the vertical and horizontal velocity component 
respectively, in agreement with 
the results shown in \cite{vc_pof09} (of course, the precise values of the
coefficients depend on the definition of $h$). 
The anisotropy of the large scale flow is
reflected in the velocity correlation length: the integral
scale based on horizontal velocity is smaller than 
the one based on vertical velocity. 

\begin{figure}[htb!]
\includegraphics[clip=true,keepaspectratio,width=12.0cm]{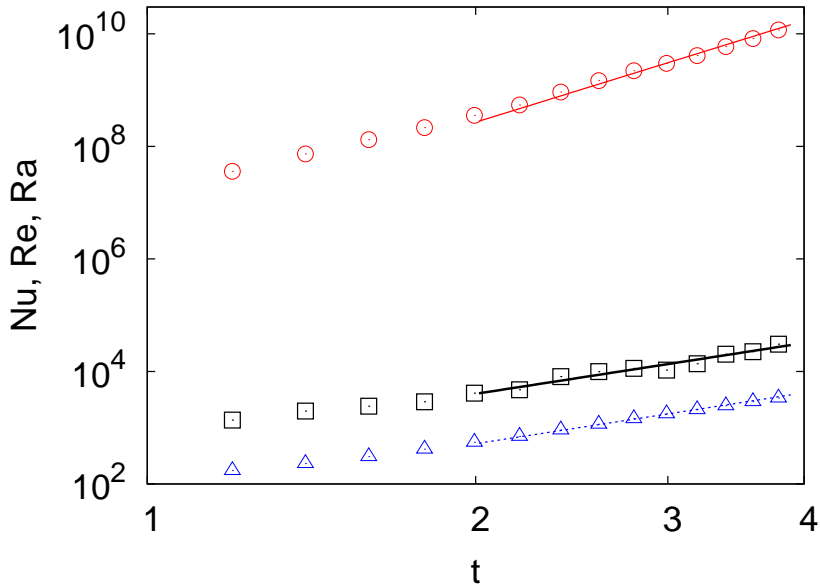}
\caption{Temporal scaling of Nusselt number 
$Nu=1+\langle w T \rangle h/(\kappa \theta_0)$ (blue triangles), 
Reynolds number
$Re=v_{rms} h/\nu$ (black squares) 
and Rayleigh number 
$Ra=\beta g \theta_0 h^3/(\nu \kappa)$
(red circles) for simulation $B$ at $Pr=1$.
The lines are the temporal scaling predictions $t^3$ for 
$Nu$ and $Re$ and $t^6$ for $Ra$.}
\label{fig3.6}
\end{figure}

\begin{figure}[htb!]
\includegraphics[clip=true,keepaspectratio,width=12.0cm]{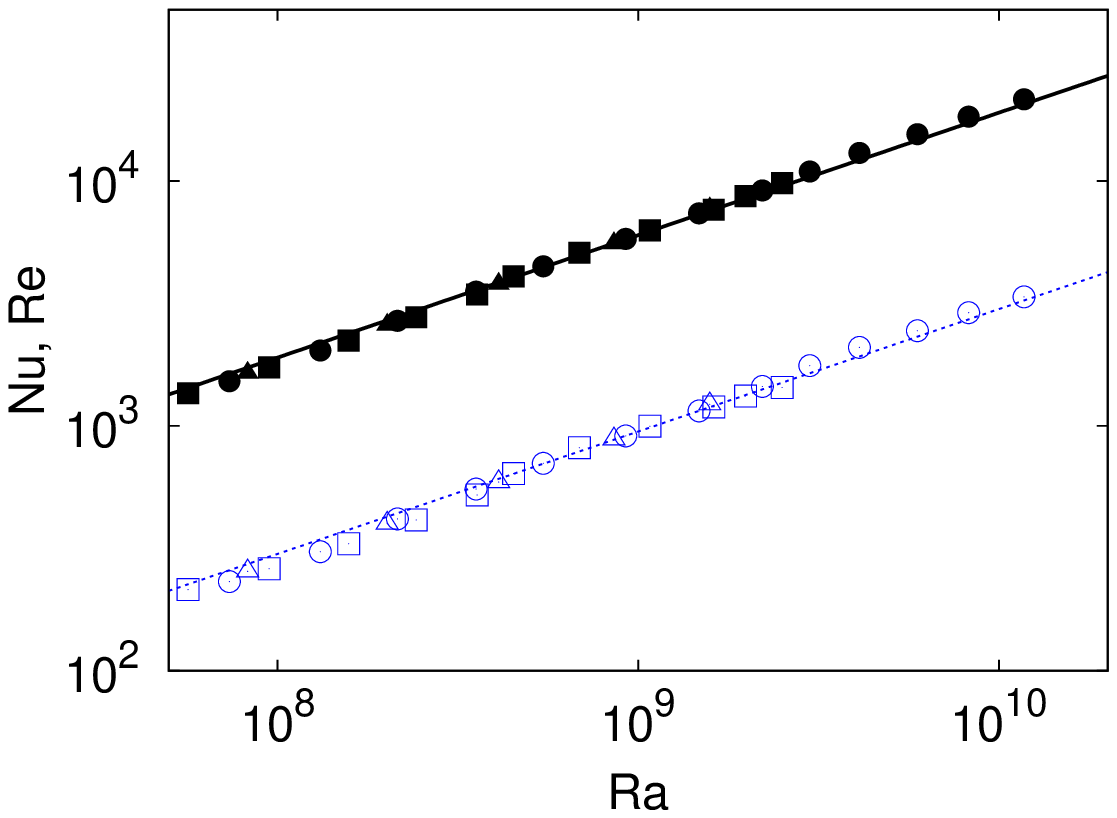}
\caption{Nusselt (blue empty symbols) and Reynolds 
(black filled symbols) numbers as a function of Rayleigh number
at $Pr=1$.
The ``ultimate state''  
prediction (black and blue lines) expressed by (\ref{eq:3})
are compared with numerical data obtained from simulations 
A (squares), B (circles) and C (triangles).}
\label{fig3.6b}
\end{figure}

We end this Section by discussing the behavior 
of the turbulent heat flux, the energy transfer and the 
mean temperature gradient in terms of dimensionless variables (as discussed 
in Sec~\ref{sec:2}): Nusselt, Reynolds and Rayleigh numbers, respectively. 
The temporal evolution of these numbers, shown in Fig.~\ref{fig3.6},
follows the dimensional predictions (\ref{eqnutemp})
for 
the temporal evolution of $\alpha$ (see Inset of Fig.~\ref{fig3.4}).  
The presence of the ``ultimate state of thermal convection'',
in the restricted case $Pr=1$, 
is also confirmed by our numerical results. 
Data obtained from simulations at various resolution
(see Fig.~\ref{fig3.6b}) are in close agreement 
with the ``ultimate state'' scalings (\ref{eq:3}). 

\section{Small-scale statistics}
\label{sec:4}

As already discussed in the introduction, the phenomenological theory
predicts that, at small-scales, RT turbulence realizes an adiabatically 
evolving Kolmogorov--Obukhov scenario of NS turbulence. 
Here adiabatic means that, because
of the scaling laws, small scales have sufficient time to adapt to 
the variations of large scales, leading to a scale-independent energy
flux. We remark that this is not the only possibility, as in two dimensions
the phenomenology is substantially different.
Unlike the 3D configuration, the 2D scenario is an example
of active scalar problem. Indeed, the buoyancy effect
is leading at both large and smaller scales.
An adiabatic generalization of Bolgiano--Obukhov scaling
has been predicted by means of mean field theory \cite{chertkov_prl03}  and
has been confirmed numerically \cite{cmv_prl06}.

\begin{figure}[htb!]
\includegraphics[clip=true,keepaspectratio,width=12.0cm]{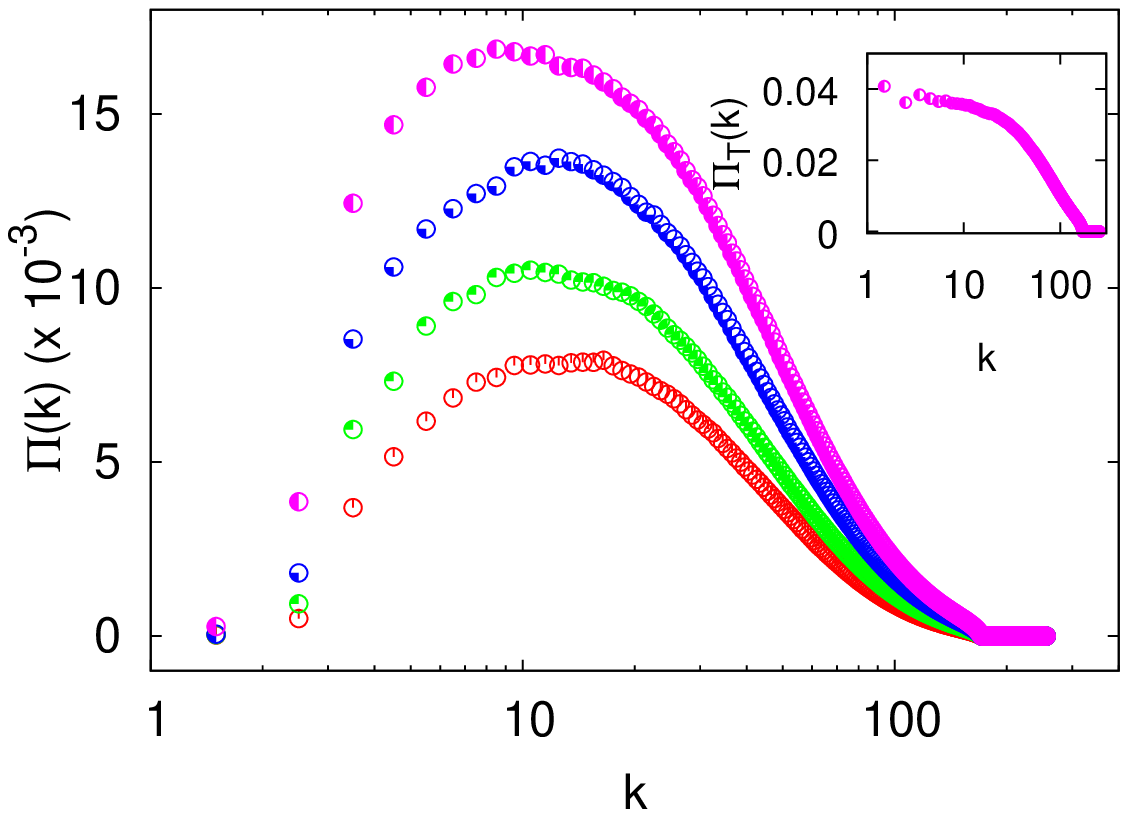}
\caption{Spectral global kinetic energy flux $\Pi(k)$ at times 
$t=2.4 \tau$, $t=2.6 \tau$, 
$t=2.8 \tau$, $t=3.0 \tau$ (from bottom to top)
and temperature variance flux at $t=3.0 \tau$ (inset)
for simulation B. Kinetic energy flux is defined as
$\Pi(k) = - \int_k^{\infty} Re \left[ \hat{v}_i(-\bf{k}')
\widehat{({\bf v}\cdot \nabla {v_i} )} ({\bf k}') \right] d{\bf k}'$
where $\hat{}\,$ is the Fourier transform \cite{frisch_95}. 
A similar definition holds for the temperature variance flux.
}
\label{fig4.1}
\end{figure}

Figure~\ref{fig4.1} shows the global energy flux in spectral space
at different times in the turbulent stage of the simulation. 
As discussed above, the flux grows in time following the increase
of the input $\mathcal{I}(t)$ at large scales and at smaller ones, faster scales have time
to adjust their intensities to generate a scale independent flux.

If the analogy with NS turbulence is taken seriously, one can extend the
dimensional predictions (\ref{eq:4}-\ref{eq:5}) to include intermittency
effects. Structure functions for velocity and temperature fluctuations
are therefore expected to follow
\begin{eqnarray}
S_{p}(r,t) &\equiv& \langle (\delta_r v_{\parallel}(t))^p \rangle 
\simeq v_{rms}(t)^p \left({r \over h(t)} \right)^{\zeta_p}
\label{eq:4.1} \\
S^{T}_{p}(r,t) &\equiv& \langle (\delta_r \theta(t))^p \rangle 
\simeq \theta_0^p \left({r \over h(t)} \right)^{\zeta^T_p}
\label{eq:4.2}
\end{eqnarray}
In (\ref{eq:4.1}) we introduce the longitudinal velocity differences
$\delta_r v_{\parallel}(t) \equiv ({\bf v}({\bf x}+{\bf r},t) 
-{\bf v}({\bf x},t)) \cdot {\bf r}/r$ and the increment $r$ is 
made dimensionless
with a characteristic large scale which, in the present setup, is
proportional to the width of the mixing layer $h(t)$, the only scale
present in the system.
The two sets of scaling exponents $\zeta_p$ and $\zeta^T_p$
are known from both experiments \cite{war_ann00,tabeling} 
and numerical simulations
\cite{wg_njp04} with good accuracy for moderate $p$. 
Mean-field prediction is $\zeta_p=\zeta^T_p=p/3$ while intermittency
leads to a deviation with respect to this linear behavior.
Kolmogorov's ``4/5'' law for third-order velocity implies
the exact result $\zeta_3=1$, while temperature exponents are not fixed, 
apart for standard inequality requirements
\cite{frisch_95}. Both experiments and simulations give stronger
intermittency in temperature than in velocity fluctuations, {\it i.e.}
$\zeta^T_p < \zeta_p$ for large $p$.

We have computed velocity and temperature structure functions
and spectra in our simulations of RT turbulence. 
To overcome the inhomogeneity of the setup, velocity and temperature
differences (at fixed time) are taken between points both belonging 
to the mixing layer as defined above.
Isotropy is recovered by averaging the separation ${\bf r}$ over 
all directions. Spectra are computed by Fourier-transforming 
velocity and temperature fields on two-dimensional horizontal
planes and then averaging vertically over the mixing layer.

\subsection{Lower-order statistics}
%

\begin{figure}[htb!]
\begin{tabular}{@{}c@{}c@{}c@{}}
{\includegraphics[clip=true,keepaspectratio,width=8.0cm]{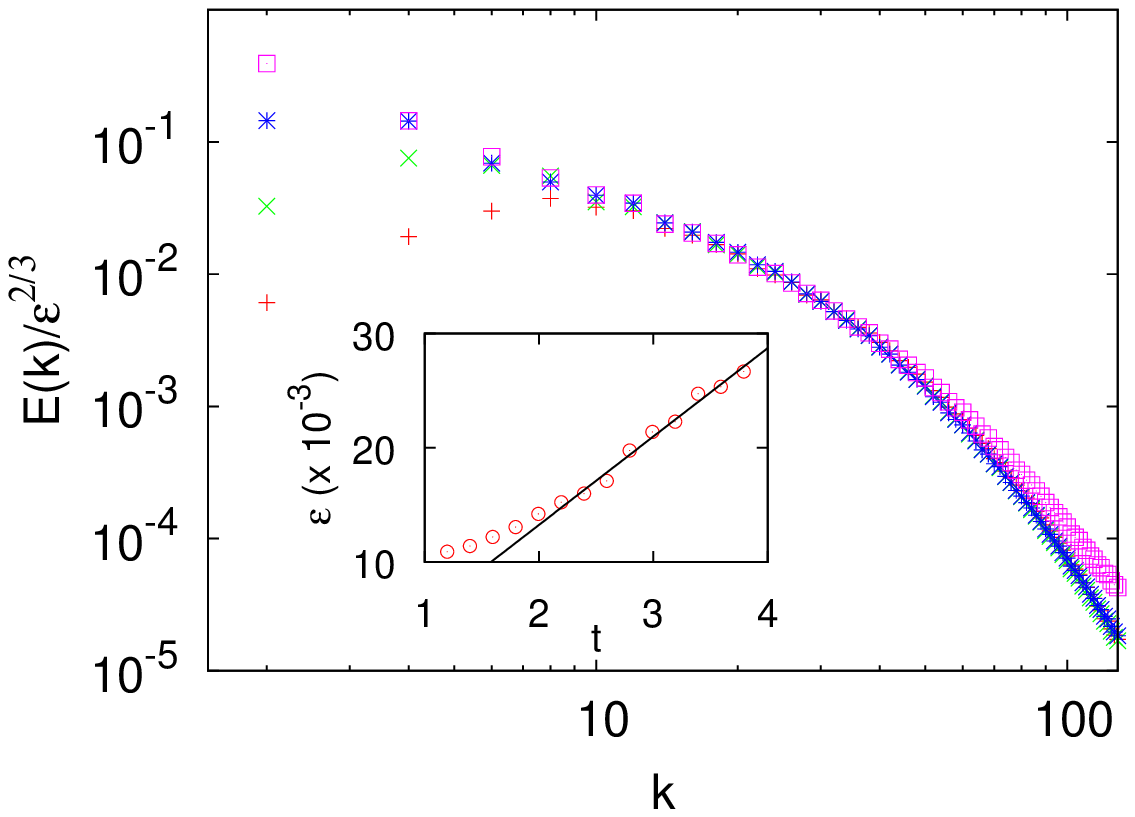}}&~~~~&
\includegraphics[clip=true,keepaspectratio,width=8.0cm]{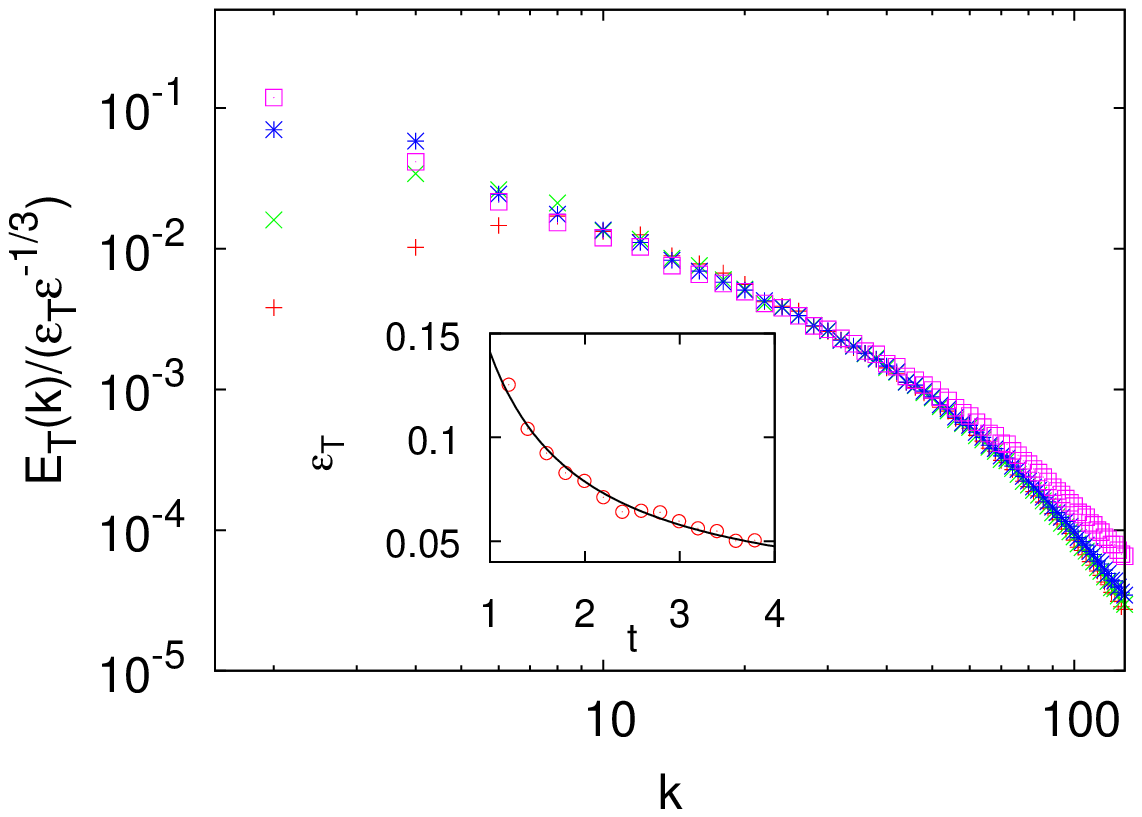}\\
{\footnotesize\textbf{(a)}}&~~~~&
{\footnotesize\textbf{(b)}}\\
\end{tabular}
\caption{\textbf{(a)} Kinetic-energy spectra compensated with $\epsilon^{2/3}$
at times $t=1 \tau$ (red crosses), 
$t=1.4 \tau$ (green  times),
$t=1.8 \tau$ (blue stars) and $t=3.8 \tau$ (pink squares).
Inset: kinetic-energy dissipation {\it vs.} time. The line represents 
the linear growing of energy dissipation (see Sec.~\ref{sec:2}). 
\textbf{(b)} Temperature-variance spectra compensated 
with $\epsilon_T^{-1}\epsilon^{1/ 3}$ at same times.   
Inset: temperature-variance dissipation {\it vs.} time. The line is 
the dimensional prediction $\sim t^{-1}$ (see Sec.~\ref{sec:2}). 
Data from simulation B.
}
\label{fig4.2}
\end{figure}

\begin{figure}[htb!]
\includegraphics[clip=true,keepaspectratio,width=12.0cm]{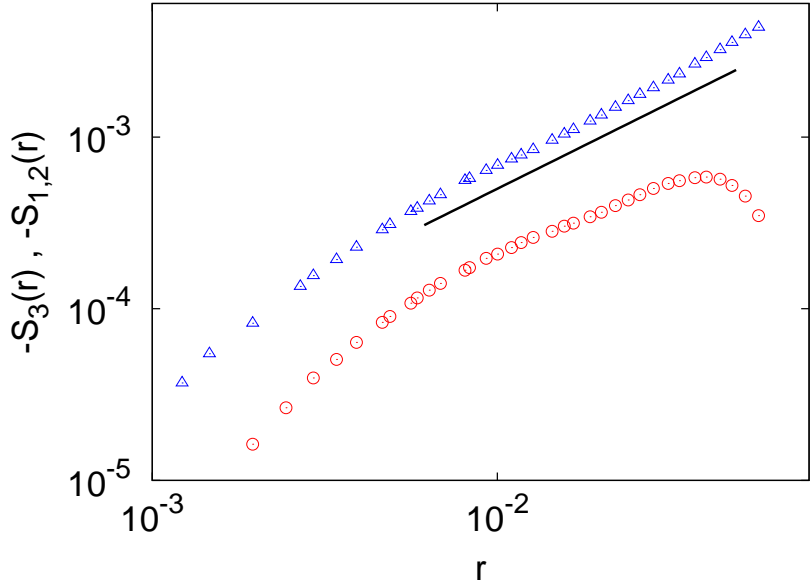}
\caption{Third-order isotropic longitudinal velocity structure
function $S_3(r)$ computed at a late stage in the simulation (red circles)
and mixed longitudinal velocity-temperature structure function 
$S_{1,2}(r)$ (blue triangles). The black line represents the linear
scaling. Data from simulation B.}
\label{fig4.3}
\end{figure}

In Fig.~\ref{fig4.2}(a), we 
plot kinetic-energy spectra at different times in the 
turbulent stage, compensated with the time 
dependent energy dissipation $\epsilon^{2/3}(t)$. In the intermediate 
range of wavenumbers, corresponding to inertial scales, the collapse is almost 
perfect. The evolution of the compensated spectra 
shows that the growth of the integral scale at small wavenumbers is
in agreement with Fig.~\ref{fig3.5}. 
Likewise temperature-variance spectra are considered in Fig.~\ref{fig4.2}(b).  
Here, the spectra are compensated with both the time dependent temperature 
variance dissipation $\epsilon_T^{-1}(t)$ and the energy 
dissipation $\epsilon^{1/3}(t)$. The  evolution of the intermediate range of 
wavenumbers follows the dimensional prediction (\ref{eq:5}). 

Figure~\ref{fig4.3} displays the third-order velocity structure function
$S_3(r)$, related to the energy flux by Kolmogorov's ``4/5'' law
$S_3(r)=-(4/5) \epsilon\,r$ \cite{frisch_95}. We also plot the mixed 
velocity-temperature structure function 
$S_{1,2}(r) \equiv \langle \delta_r v_{\parallel} (\delta_r T)^2 \rangle$
which is proportional to the (constant) flux of temperature fluctuations 
$\epsilon_T$ according to Yaglom's law 
$S_{1,2}(r)=-(4/3) \epsilon_T\,r$ \cite{Yag_dan49}. 
Both the computed structure functions
display a range of linear scaling, {\it i.e.} a constant flux, in the inertial 
range of scales $5 \times 10^{-3} \le r/L_z \le 5 \times 10^{-2}$. 
It is interesting to
observe that the mixed structure function $S_{1,2}(r)$ seems to have a 
range of scaling which extends to larger scales.
This is probably due to the fact that at large-scale temperature
fluctuations are dominated by unmixed plumes which have strong correlations
with vertical velocity.

\begin{figure}[htb!]
\includegraphics[clip=true,keepaspectratio,width=12.0cm]{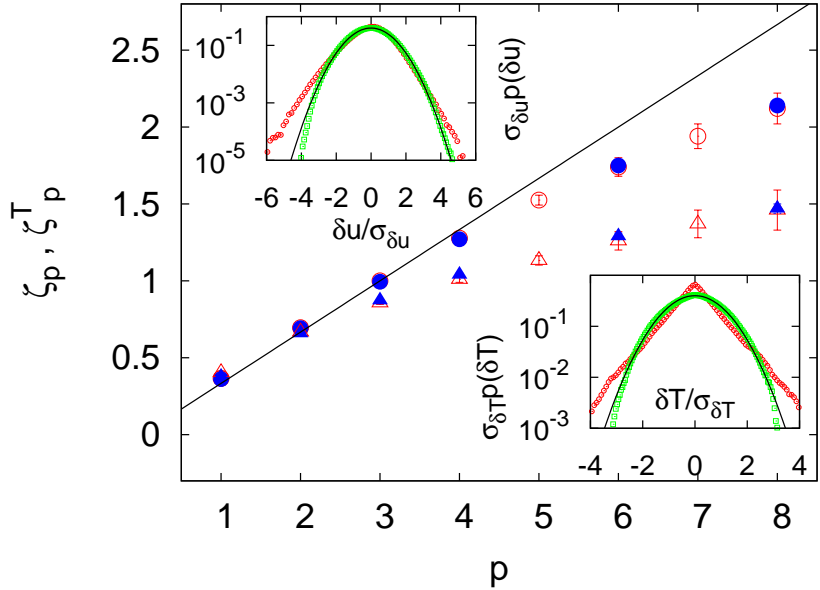}
\caption{Structure-function scaling exponents for velocity increments
$\zeta_p$ (circles) and temperature increments $\zeta^{T}_p$ (triangles)
with absolute values.
Red open symbols are obtained using ESS procedure \cite{ESS} on the present
simulation at time $t=3\,\tau$, fixing the value of $\zeta_3=1$ and 
$\zeta^{T}_2=2/3$. Errors represent fluctuations observed in 
different realizations of simulation B.
Blue filled symbols are taken from a stationary NS simulation 
at $R_{\lambda}=427$ \cite{wg_njp04}. Black line is Kolmogorov
non-intermittent scaling $p/3$.
Insets: probability density function for velocity differences 
$\delta_r v(t)$ (upper) and temperature differences $\delta_r T(t)$
at time $t=3 \tau$ and scales $r=0.008 L_z$ (red circles)
and $r=0.06 L_z$ (green squares). Black lines represent a standard
Gaussian.}
\label{fig4.4}
\end{figure}

%
\subsection{Spatial/temporal intermittency}

Despite the clear scaling observable in Fig.~\ref{fig4.3}, it is very
difficult to compute scaling exponents directly from
higher-order structure functions because of limited Reynolds number
and statistics. Therefore, assuming a scaling region as in Fig.~\ref{fig4.3},
we can compute {\it relative} scaling exponents using the so-called
Extended Self Similarity procedure \cite{ESS}. This corresponds to
consider the scaling of one structure function with respect to a reference
one ({\it e.g.} $S_3(r)$ for velocity statistics), and thus to measure 
a relative exponent ({\it i.e.} $\zeta_p/\zeta_3$). 

Scaling exponents obtained in this way are shown in Fig.~\ref{fig4.4}.
Reference exponents for the ESS procedure are $\zeta_3=1$ and $\zeta^T_2=2/3$
(which is not an exact result). We see that both velocity and temperature
scaling exponents deviate from the dimensional prediction of 
(\ref{eq:4}-\ref{eq:5}) ({\it i.e.} $\zeta_p=\zeta^T_p=p/3$) indicating
intermittency in the inertial range. We also observe a stronger
deviation for temperature exponents, which is consistent with what
is known for the statistics of a passive scalar advected by a 
turbulent flow \cite{frisch_95,sa_arfm97}.

The question regarding the universality of the set of 
scaling exponents with respect to the geometry and the large-scale
forcing naturally arises.
Several experimental and numerical investigations in three-dimensional
turbulence support the universality scenario in which the set of
velocity and passive-scalar scaling exponents are independent of 
the details of large-scale energy injection and geometry of the flow.
Therefore, because we have seen that in 3D RT turbulence at small 
scales temperature becomes passively transported and isotropy is recovered,
one is tempted to compare scaling exponents with those obtained in
NS turbulence.
As shown in Fig.~\ref{fig4.4},
the two sets of exponents coincide, within the error bars, with
the exponents obtained from a standard NS simulation with passive
scalar at comparable $R_{\lambda}$ \cite{wg_njp04}.

We remark that scaling exponents for passive scalar in 
NS turbulence are very sensitive to the fitting procedure. 
Strong temporal fluctuations
have been observed in single realization \cite{cc_prl97} and dependence
on the fitting region has been reported \cite{wg_njp04}. 
Indeed, different realizations of RT turbulence (starting with
slightly different initial perturbations) lead to fluctuations of
scaling exponents which account for the errorbars shown in 
Fig.~\ref{fig4.4}.

Figure~\ref{fig4.4} also shows probability density functions for velocity
and temperature fluctuations for two different scales. 
Both distributions are close to a Gaussian at large scale and develop
wide tails at small scales, indicating the absence of self-similarity
thus confirming the intermittency scenario.

\begin{figure}[htb!]
\includegraphics[clip=true,keepaspectratio,width=12.0cm]{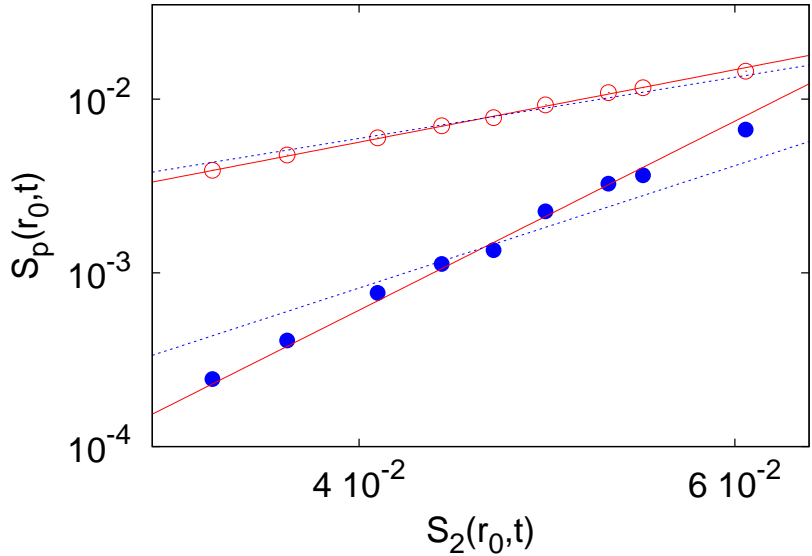}
\caption{Time dependence of $p$-order velocity structure function
$S_{p}(r_0,t)$ {\it vs.} $S_{2}(r_0,t)$ for $p=4$ (red open circles) and 
$p=8$ (blue filled circles) with $r_0/L_z=0.012$, in the 
middle of the inertial range for simulation B. 
Red, continuous lines represent 
the intermittent prediction $\beta_p=p-2 \zeta_p$ with
$\zeta_p$ given by spatial structure functions; blue dashed lines
are the non-intermittent prediction $\beta_p=p/3$.}
\label{fig4.5}
\end{figure}

As a further numerical support of (\ref{eq:4.1}-\ref{eq:4.2}) 
we now consider temporal
behavior of structure functions. From (\ref{eq:4.1}), taking into 
account the temporal evolution of large scale quantities,
we expect the temporal scaling 
$S_{p}(t)\sim t^{\beta_p}$ with $\beta_p=p-2 \zeta_p$.
With Kolmogorov scaling one simply has $\beta_p=p/3$ but intermittent
corrections are expected to be important, for example 
$\beta_6 \simeq 2.4$ instead of $p/3=2$. 
Figure~\ref{fig4.5}
shows the scaling of $S_{p}(r,t)$ {\it vs.} $S_{2}(r,t)$ ({\it i.e.} in the
ESS framework) for a particular value of $r=r_0=0.0012 L_z$. 
The relative temporal exponents $\beta_p/\beta_2$
obtained from the spatial exponents $\zeta_p$ of Fig.~\ref{fig4.4} 
fit well the data, while non-intermittent relative scaling exponents
$\beta_p/\beta_2=p/2$ are ruled out.

The effects of intermittency are particularly important at very
small scales. One important example is the statistics of 
acceleration which has recently been the object of experimental
and numerical investigations \cite{LaPorta_acc, bbcdlt_prl04}. 
For completeness, we briefly recall the
main results obtained in those studies. 

\begin{figure}[htb!]
\includegraphics[clip=true,keepaspectratio,width=12.0cm]{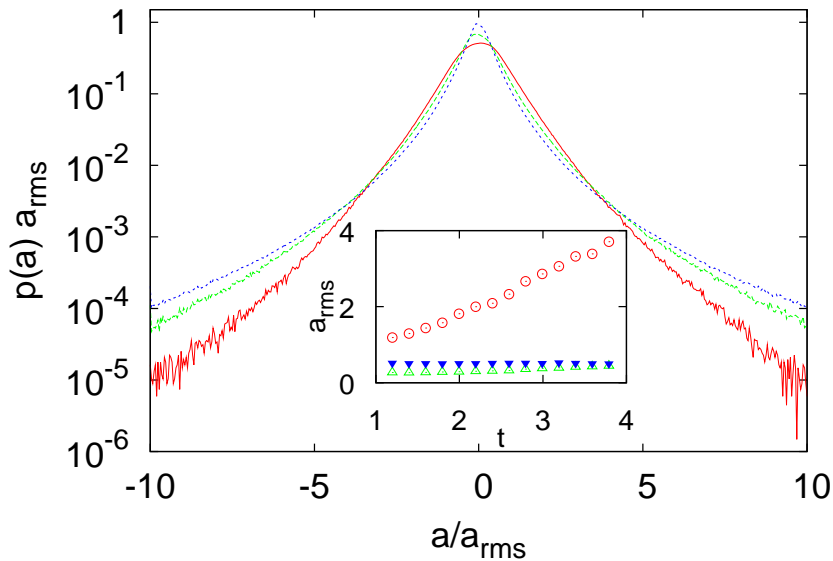}
\caption{Probability density function of the vertical component 
of the acceleration at time $t= 1\,\tau$ (red, bottom
tails), $t=2\,\tau$ (green, intermediate tails) and $t=3.8\,\tau$ (blue, 
upper tails) normalized with rms values. Inset: evolution of $a_{rms}$
with time for the three contributions of (\ref{eq:1}): 
pressure gradient $\partial_z p$
(red circles), dissipation $\nu \triangle w$ (green open triangles) 
and buoyancy term $\beta g T$ (blue filled triangles). Data from 
simulation B.
}
\label{fig4.6}
\end{figure}

The acceleration $a$ of a Lagrangian particle transported by the
turbulent flow is by definition given by the r.h.s of (\ref{eq:1}).
In the present case of Boussinesq approximation, the acceleration
has three contributions: pressure gradient, viscous dissipation and
buoyancy terms. Neglecting intermittency for the moment, dimensional
scaling (\ref{eq:4}-\ref{eq:5}) implies that 
$-\nabla p \simeq \nu \triangle u \simeq \nu^{-1/4} 
(\beta g \theta_0)^{3/2} t^{3/4}$ while $\beta g T \simeq \beta g \theta_0$.
Therefore the buoyancy term in (\ref{eq:1}) becomes subleading 
not only going to small scales but also at later times. 
Among the other two terms, 
we find that, as in standard NS turbulence, the pressure gradient
term is by far the dominant one, as shown in the inset of Fig.~\ref{fig4.6}. 
After an initial transient, we have that for $t \ge 2 \,\tau$ both terms grow
with a constant ratio $(\partial_z p)_{rms}/(\nu \triangle w)_{rms} \simeq 8$. 

The inset of Fig.~\ref{fig4.6} suggests that the temporal growth 
of $a_{rms}$ is faster than $t^{3/4}$. Again, this can be understood
as an effect of intermittency which is particularly important at 
small scales. Indeed, using the multifractal model of intermittency 
\cite{frisch_95} one obtains the prediction
$a_{rms} \sim t^{0.86}$ \cite{bbcdlt_prl04}.

The effect of intermittency on acceleration statistics 
is evident by looking at the probability density function. 
Figure~\ref{fig4.6} shows that the distribution develops 
larger tails as turbulence intensity,
and Reynolds number, increases. 
This effect is indeed expected, as the shape of the acceleration pdf 
depends on the Reynolds number and therefore no universal form 
is reached. Nevertheless, given the value of $R_{\lambda}$ as a
parameter, the pdf can be predicted again using the multifractal
model \cite{bbcdlt_prl04}.

\section{Conclusion}
\label{sec:5}
We have studied spatial and temporal statistics of Rayleigh--Taylor
turbulence in three dimensions 
at small Atwood number and at Prandtl number one 
on the basis of a set of
high resolution numerical simulations.
RT turbulence is a paradigmatic example of non-stationary
turbulence with a time dependent injection scale. 
The phenomenological theory proposed by Chertkov \cite{chertkov_prl03}
is based on the notion of adiabaticity where 
small scales are slaved to large ones: the latter are forced by
conversion of potential energy into kinetic energy; the former undergo
a turbulence cascade flowing to smaller scales
until molecular viscosity becomes important.
In this picture, temperature actively forces hydrodynamic degrees of 
freedom at large scales while it behaves like
a passive scalar field at small scales where a constant kinetic energy
flux develops.

The above scenario suggests comparison of RT turbulence with 
classical homogeneous, isotropic, stationary 
Navier--Stokes turbulence, in the general framework of the
existence of universality classes in turbulence.

By means of accurate direct numerical simulations, we provide
numerical evidence in favor of the mean-field theory.
Moreover, we extend the analysis to higher order 
statistics thus addressing the issue related to intermittency 
corrections.
By measuring scaling exponents of both velocity and temperature
structure functions, we find that indeed they are compatible with
those obtained in standard turbulence. This result gives further
support for the universality scenario.

We also investigate temporal evolution of global quantities, 
both geometrical (the width of mixing layer) and dynamical 
(the heat flux). The relevant dimensionless quantity in RT
turbulence are the Rayleigh, Reynolds and Nusselt numbers for which
there exists an old prediction due to Kraichnan \cite{kraichnan_pof62}, 
known as the ``ultimate state of thermal convection'', which links the 
dimensionless number in terms of simple scaling laws.
Our set of numerical simulations give again strong evidence 
for the validity of such scaling in RT turbulence 
at small Atwood number and at Prandtl number one 
thus confirming
how important in thermal convection is the role of boundaries which 
prevent the emergence of the ultimate state.


\bibliography{biblio}{}

\end{document}